\documentclass[a4paper,twoside]{article}
\usepackage{jadt2004}
\usepackage{amsmath,amsthm,amssymb}
\usepackage[francais]{babel}
\usepackage[latin1]{inputenc}
\usepackage[dvips]{graphics}
\usepackage[T1]{fontenc}
\usepackage{ae}

\begin{document}

\title{Analyse et expansion des textes en question-réponse}

\author{Bernard Jacquemin}

\email{Bernard.Jacquemin@xrce.xerox.com}

\affils{ILPGA - 19, rue des Bernardins - 75\,005 Paris - France \\
        XRCE - 6, chemin de Maupertuis - 38\,240 Meylan - France}
\maketitle

\JADTEvenHeader{Bernard Jacquemin}

\JADTOddHeader{Analyse et expansion des textes en
question-réponse}

\begin{Abstract}
This paper presents an original methodology to consider question
answering. We noticed that query expansion is often incorrect
because of a bad understanding of the question. But the automatic
good understanding of an utterance is linked to the context
length, and the question are often short. This methodology
proposes to analyse the documents and to construct an informative
structure from the results of the analysis and from a semantic
text expansion. The linguistic analysis identifies words
(tokenization and morphological analysis), links between words
(syntactic analysis) and word sense (semantic disambiguation). The
text expansion adds to each word the synonyms matching its sense
and replaces the words in the utterances by derivatives, modifying
the syntactic schema if necessary. In this way, whatever
enrichment may be, the text keeps the same meaning, but each piece
of information matches many realisations. The questioning method
consists in constructing a local informative structure without
enrichment, and matches it with the documentary structure. If a
sentence in the informative structure matches the question
structure, this sentence is the answer to the question.
\end{Abstract}

\begin{Resume}
Cet article présente une méthode originale d'envisager la tâche de
question-réponse. Nous avons remarqué que l'expansion de requête
est souvent erronée du fait d'une mécompréhension de la question.
Mais la bonne compréhension d'un énoncé est fonction de la taille
du contexte, et les question sont souvent courtes. Notre approche
propose d'analyser les documents et de construire une structure
informationnelle en utilisant les résultats de l'analyse ainsi que
l'enrichissement sémantique des textes. Une analyse linguistique
identifie les mots (segmentation et analyse morphologique), les
liens entre les mots (analyse syntaxique) et le sens des mots
(désambiguïsation sémantique lexicale). L'expansion de texte
adjoint à chaque mot les synonymes qui correspondent à son
acception contextuelle et remplacent les mots des énoncés par
leurs dérivés, modifiant si nécessaire la structure syntaxique de
la phrase. De la sorte, quel que soit l'enrichissement utilisé, le
texte conserve la même signification, tandis que chaque élément
d'information obtient de nombreuses actualisations. La procédure
d'interrogation consiste à construire une structure
informationnelle locale à la question sans effectuer
d'enrichissement, et à faire lui faire correspondre un ou
plusieurs fragments de la structure documentaire. Si une phrase de
la structure informationnelle documentaire correspond à celle de
la question, cette phrase contient la réponse à la question.
\end{Resume}


\section{Introduction}

Dans notre société, la maîtrise de l'information est devenue un
enjeu très important, que ce soit en politique, en économie, dans
le domaine culturel, etc. Mais le contrôle de l'information doit
faire face à la profusion des données disponibles. De fait, le
nombre des documents textuels augmente chaque jour, et aucun
humain ne peut actuellement avoir une idée précise de l'ensemble
de l'information qu'ils contiennent.

De ce fait, les méthodes qui permettent d'étudier, d'identifier et
de classifier l'information dans les textes électroniques prennent
de plus en plus d'importance. Les disciplines de traitement de
l'information - et plus particulièrement l'extraction
d'information, la recherche d'information ou la tâche de
question-réponse - permettent de trouver un élément d'information
dans une base textuelle. Dans ce domaine, chacune des disciplines
possède ses propres spécificités. En particulier, la tâche de
question-réponse joue un rôle de premier plan d'une part parce que
contrairement à la recherche d'information, c'est une information
particulière et précise qui doit être obtenue, et d'autre part
parce que les données recherchées sont variable, contrairement à
la discipline d'extraction d'information.

Depuis TREC-8 (\textsl{Text REtrieval Conference})
\cite{VoorheesHarman99,Voorhees99}, toutes les méthodes de
question-réponse proposées utilisent une architecture similaire :
un module d'analyse de la question chargé d'identifier les
éléments d'information présents dans la requête, et parfois de
catégoriser la réponse attendue ; un système d'expansion de
requêtes qui donne à chaque élément d'information de la question
autant d'actualisations qu'il est possible, car la même donnée
peut se présenter sous plusieurs formes dans un texte ; un moteur
de recherche qui recherche chaque élément d'information dans les
textes, quelle que soit son actualisation.

Toutes les méthodes qui participent à la compétition TREC fondent
leur méthode de recherche sur le traitement de la requête. Par
exemple, \cite{Hull99} utilise des analyseurs pour extraire et
identifier le vocabulaire de la question et trouver la réponse
contenant les mots les plus importants de cette question. Le
système QALC \cite{FerretAl02} est basé sur une analyse
linguistique de la requête, ainsi que sur l'exploitation par un
moteur de recherche de mots-clefs obtenus au départ de cette
requête (racinisation, synonymes, entités nommées etc.). Falcon
\cite{HarabagiuAl02} est le seul système qui considère que le
traitement des données textuelles est aussi important que la
catégorisation des questions. Cependant, ce système est également
centré sur le traitement de la requête. En effet, les fragments de
la base textuelle analysés sont obtenus par un moteur de recherche
utilisant le vocabulaire de la question ou des mots-clefs issus du
traitement de cette question.

Nous avons décidé de faire varier cette architecture
traditionnelle en effectuant les principaux traitements non plus
sur la requête, mais sur les textes de la base documentaire dans
laquelle sont cherchées les réponses. Au travers de ces
traitements, nous obtenons une structure informationnelle enrichie
que nous pouvons interroger par un traitement minimal de la
question. Dans cet article, nous exposons d'abord pourquoi nous
avons choisi d'effectuer des traitements sur les textes. Ensuite,
nous présentons le traitement appliqué aux documents pour récolter
des données destinées à la construction de la structure
informationnelle. Puis nous décrivons brièvement la technique
d'interrogation de la structure. Enfin, nous présentons un
évaluation de ce système de question-réponse.

\section{Effectuer des traitements sur les textes plutôt que sur les
questions}

La décision de procéder à des traitement d'expansion sur les
documents plutôt que sur la question provient d'une constatation :
de nombreux enrichissements appliqués à la question ne
correspondent pas à la signification de l'élément d'information
qu'ils enrichissent dans le contexte qu'il occupe. Dans l'exemple
\ref{expSyn}, on voit que la méconnaissance du sens d'une unité
lexicale conduit à effectuer une expansion erronnée dans deux cas
 (\textit{universel} et \textit{vague}) pour une expansion correcte dans un seul
 cas (\textit{commandant}). Or si le sens correct de l'unité
 lexicale ainsi enrichie par synonymie était connu, l'expansion
 serait correcte et des erreurs pourraient être évitées.

\begin{figure}[!h]
\begin{center}
\mbox{\parbox{10cm}{
\textsf{Question proposée~:}\\
Qui est le général des Perses~?\\

\textsf{Synonymes de général~:}\\
commandant (sens n°2)\\
universel (sens n°1)\\
vague (sens n°3)\\

\textsf{Question enrichie par tous les synonymes~:}\\
$\mathrm{Qui~est~le~}
\begin{array}{l}
\mathrm{g\acute{e}n\acute{e}ral}  \\
\mathrm{commandant} \\
\mathrm{universel}  \\
\mathrm{vague} \\
\end{array}
\mathrm{~des~Perses~?} $\\

\textsf{Question enrichie par les synonymes propres au sens ~:}\\
$\mathrm{Qui~est~le~}
\begin{array}{l}
\mathrm{g\acute{e}n\acute{e}ral}  \\
\mathrm{commandant} \\
\end{array}
\mathrm{~des~Perses~?} $\\ }}
 \caption{Expansion synonymique~:
intérêt de la sémantique.} \label{expSyn}
\end{center}
\end{figure}

Nous avons tiré les mêmes conclusions pour les autres types
d'enrichissements : racinisation, utilisation de dérivés, etc.
Chacune de ces méthodes d'expansion rencontre le problème de la
bonne compréhension de la signification des éléments d'information
de la question.

Or lorsque l'expansion de la requête apporte des enrichissements
qui ne correspondent pas à la signification des éléments qui
constituent cette requête, ces enrichissements erronés sont
susceptibles d'amener des réponses incorrectes. Cependant,
l'utilisation d'un système capable de déterminer le sens exact de
chaque élément composant la question - c'est-à-dire un système de
désambiguïsation sémantique - ne semble pas capable de résoudre le
problème. En effet, ces systèmes peuvent déterminer le sens des
unités lexicales dans un énoncé par l'analyse du contexte,
notamment lexical et syntaxique. Le contexte des questions est
généralement court, et de plus les questions utilisent le plus
souvent une syntaxe qui leur est particulière. Comme les documents
permettent le plus souvent de disposer d'un contexte plus large
que les questions, et que leur syntaxe est généralement plus
conforme aux possibilités des grammaires des analyseurs
syntaxiques, nous avons jugé qu'ils étaient plus propices à
l'analyse sémantique, et donc à l'expansion d'énoncés.

Nous avons par ailleurs noté que les procédures appliquées aux
questions étaient lourdes et contraignantes. De plus, au plus le
système est subtil et les résultats de qualité, au plus les
traitements appliqué aux questions sont substantiels. Mais
l'importance de ces traitements est inversement proportionnelle à
l'utilisabilité du système. Cependant, si les traitements les plus
lourds sont appliqués aux documents avant l'utilisation du système
plutôt qu'à la question lors de l'interrogation, la réponse à
cette question peut être presque instantanée. Pour ces raisons,
nous avons choisi d'appliquer les traitements aux documents de la
base textuelle, d'enrichir ces documents et de conserver un index
tant des données ajoutées que des informations originelles.

\section{Traitements appliqués aux documents}

Notre méthodologie est centrés sur des techniques linguistiques et
sur l'utilisation de ressources lexicales. Nous considérons que
l'information présente dans les textes est exprimée par les mots
<<~significatifs~>>, qui sont les éléments d'information, par les
relations syntaxiques qui relient ces éléments entre eux, et par
le sens que ces mots possèdent dans le contexte où ils
apparaissent, qui correspond à la signification des éléments
d'information. De ce fait, l'analyse des textes consiste en une
identification des unités lexicales qui les composent, des
relations syntaxiques qui y apparaissent et du sens des mots.

Pour l'analyse morphologique et l'extraction des liens syntaxiques
entre les unités lexicales, nous utilisons des outils conçus au
\textsl{Xerox Research Centre Europe} (XRCE) de Grenoble.
L'analyseur morphologique s'appelle NTM (\textsl{Normalizer -
Tokenizer - Morphological analyser}) \cite{Ait98,Trouilleux01}.
Cet analyseur met en concurrence plusieurs transducteurs qui
contiennent le lexique de la langue, des règles de génération qui
constituent le <<~devin~>> (\textsl{guesser}), des règles de
flexion morphologique, etc. Il permet d'obtenir la segmentation du
texte en unités lexicales et propose pour chacune toutes les
analyses morphologiques possibles. Par ailleurs, toute information
ajoutée au transducteur lexical peut se retrouver comme une
étiquette supplémentaire dans les analyses proposées.

L'analyseur syntaxique mis en {\oe}uvre est XIP (\textsl{Xerox
Incremental Parser}) \cite{Roux99, AitAl02, HagegeRoux02}. Cet
analyseur robuste crée des syntagmes minimaux (\textsl{chunks}) et
extrait les dépendances syntaxiques qui existent entre la tête des
différents syntagmes minimaux. Pour effectuer ces opérations, il
exploite des règles grammaticales. Les règles exploitent le
contexte : catégorie grammaticale des unités lexicales, nature des
syntagmes minimaux (nominaux, verbaux, prépositionnels, etc.) ou
autres traits. Les grammaires XIP sont très souples, et cette
flexibilité permet d'ajouter de nouvelles règles, de nouveaux
types de dépendances, de nouveaux traits portant soit sur des
mots, soit sur des dépendances ou encore sur des syntagmes
minimaux, selon les besoins. En particulier, il est possible
d'affecter des traits sémantiques sur des unités lexicales et d'en
confier la gestion à une grammaire spécifique.

Le désambiguïsateur sémantique que nous utilisons est une
évolution du système développé naguère à XRCE
\cite{SegondAl00,BrunAl01}. Ce système exploitait les données
extraite des exemples et des collocateurs du dictionnaire bilingue
\textit{Oxford Hachette French Dictionary} \cite{CorreardGrundy94}
comme une information lexico-syntaxique de désambiguïsation. Le
nouveau système de désambiguïsation sémantique
\cite{JacqueminAl02,Jacquemin03} fonctionne toujours sur base de
règles de dés\-am\-bi\-guï\-sa\-tion obtenues à partir de
l'information d'un dictionnaire, le \textit{Dictionnaire des
verbes et des mots français} (\textit{Dubois})
\cite{DuboisDuboisCharlier97}. Le système construit d'abord les
règles de désambiguïsation sémantique, exploitant pour ce faire
l'information lexicale et syntaxique contenue dans le dictionnaire
(\textsl{cf.} exemple \ref{desambRule} page \pageref{desambRule}).
Les règles de désambiguïsation sémantique sont créées selon le
formalisme des grammaires XIP. Elles proviennent de schémas
syntaxiques ou d'exemples qui sont attribués par sens aux entrées
du dictionnaire.

\begin{figure}[!h]
\begin{center}
\mbox{\parbox{12cm}{
\textsf{Exemple extrait du Dubois~:}\\
On remporte la victoire sur ses adversaires (sens n°2 : <<~gagner~>>)\\

\textsf{Une dépendance impliquant remporter~:}\\
\texttt{VARG[DIR](remporter,victoire)}\\

\textsf{Règle conditionnelle de désambiguïsation~:}\\
\texttt{remporter : VARG[DIR](remporter,victoire) ==> sens <<~gagner~>>}\\
}}
\caption{Exemple d'extraction d'une règle de désambiguïsation
sémantique.} \label{desambRule}
\end{center}
\end{figure}

L'exemple \ref{desambRule} page \pageref{desambRule} présente une
règle de désambiguïsation pour le sens n°2 <<~gagner~>> du verbe
\textit{remporter}. Cette règle permet de déterminer que dans un
contexte où l'objet direct (\texttt{VARG[DIR]}) de
\textit{remporter} est \textit{victoire}, le sens du verbe
\textit{remporter} est \textit{gagner}, et pas \textit{ramener ce
que l'on avait apporté} (sens n°1).

Donc l'application de la désambiguïsation sémantique est effectuée
par l'analyseur XIP et une grammaire composée des règles de
désambiguïsation sémantique dès lors que les dépendances
syntaxiques ont été extraites. L'application de ces règles de
désambiguïsation assigne à chaque unité lexicale désambiguïsée un
trait correspondant au numéro de sens qui lui est propre dans le
contexte. De plus, les traits sémantiques (domaine d'application,
classe sémantique) correspondant à ce sens dans le dictionnaire
sont associés à cette unité lexicale.

Le résultat de l'analyse morpho-syntaxique et de la
désambiguïsation sémantique sont stockés dans un index afin
d'obtenir une structure informationnelle des textes traités.
Chaque dépendance, chaque mot significatif, chaque trait est
conservé dans l'index et permet d'aboutir aux fragments de textes
où ils apparaissent dans la base documentaire.

\section{Enrichissement de la structure informationnelle}

En plus des traits sémantiques, c'est-à-dire le numéro de sens, la
classe sémantique et le domaine d'application provenant du
dictionnaire, qui ont été adjoints aux unités lexicales
désambiguïsées dans la structure informationnelle, d'autres
méthodes d'enrichissement ont été réalisées pour fournir autant
d'actualisation que possible aux éléments d'information. Par cette
méthode, les données présentes dans les questions pourront
coïncider avec les mêmes données présentes sous une autre forme
dans les documents.

La première méthode d'expansion des textes consiste à y introduire
des synonymes qui proviennent d'un ou plusieurs dictionnaires de
synonymes. Pour ce faire, nous utilisons les <<~parasynonymes~>>
fournis par le \textit{Dubois}, qui sont attachés au sens désigné
par le désambiguïsateur sémantique. Mais ces parasynonymes ne sont
pas assez nombreux : il y a au maximum deux synonymes par sens
dans ce dictionnaire. Comme nous disposions d'autres
dictionnaires, nous avons pu utiliser les synonymes qu'ils
fournissaient pour enrichir les textes.

Ces dictionnaires sont d'une part un dictionnaire de synonymes (le
\textit{Dictionnaire des synonymes} de René \textit{Bailly}
\cite{Bailly47}), d'autre part des dictionnaires sémantiques
(\textit{Memodata}  et \textit{EuroWordNet} français). Le problème
est que les synonymes ne sont pas distribués selon les sens du
\textit{Dubois} dans ces ressources, et que l'enrichissement ne
peut donc se faire selon la signification des unités lexicales
originelles déterminée par la désambiguïsation sémantique. Pour
pouvoir conserver un enrichissement sémantique, seuls les
synonymes qui possèdent les mêmes traits sémantiques dans le
\textit{Dubois} que le mot désambiguïsé peuvent être utilisés pour
enrichir ce mot. Les autres synonymes proposés sont considérés
comme des synonymes pour un autre sens.

La méthode d'enrichissement consiste simplement à mettre en
disjonction, dans la structure de l'information que l'on est en
train de construire, l'unité lexicale originelle et le ou les
synonymes qui lui correspondent dans l'énoncé. L'exemple
\ref{enriSyn} page \pageref{enriSyn} permet de comprendre le
fonctionnement de cet enrichissement. La désambiguïsation
sémantique a déterminé que le sens de \textit{chef} dans l'énoncé
est <<~autorité~>> (n°1). Les synonymes proposés pour ce sens (en
gras) viennent simplement se glisser en disjonction dans les
dépendance où le mot originel apparaît.

\begin{figure}[!h]
\begin{center}
\mbox{\parbox{12cm}{
\textsf{\'{E}noncé à enrichir~:}\\
César fixe à Alésia le \textbf{chef} des coalisés\\

\textsf{\textsf{Synonymes proposés~:}}\\
\textbf{commandant (sens 1~: autorité)}\\
\textbf{dirigeant (sens 1~: autorité)}\\
cuisinier (sens 2~: responsable de cuisine)\\
maître queux (sens 2~: responsable de cuisine)\\

\textsf{Dépendances impliquant chef (sens 1, structure non enrichie)~:}\\
\texttt{VARG[DIR](fixe,chef)}\\
\texttt{NMOD[INDIR](chef,de,coalisés)}\\

\textsf{Dépendances impliquant chef (sens 1, structure enrichie)~:}\\
$
\mathtt{VARG[DIR]}
\left(
\begin{array}{cc}
               & \mathtt{chef} \\
               & OU \\
\mathtt{fixe,~} & \mathtt{\mathbf{commandant}} \\
               & OU \\
               & \mathbf{\mathbf{dirigeant}} \\
\end{array} \right)
$ \\
$
\mathtt{NMOD[INDIR]}
\left(
\begin{array}{ccc}
\mathtt{chef}                & \\
OU                           & \\
\mathtt{\mathbf{commandant}} & \mathtt{,~des,~coalis\acute{e}s} \\
OU                           & \\
\mathbf{\mathbf{dirigeant}}  & \\
\end{array}
\right) $ }} \caption{Enrichissement synonymique.} \label{enriSyn}
\end{center}
\end{figure}

L'autre manière d'enrichir les documents consiste à utiliser des
formes dérivées des unités lexicales qui composent les énoncés,
considérant que la forme dérivée a un sens proche de la forme
originelle. Le \textit{Dubois} possède un champ informationnel qui
indique quels sont les dérivés d'une unité lexicale dans chacun de
ses sens. Par exemple, le dérivé \textit{tracteur} provient du mot
\textit{tirer}. Mais si \textit{tracteur} dérive bien de
\textit{tirer} sous son sens <<~remorquer~>>, il n'en va pas de
même sous son sens <<~faire partir une arme à feu~>>.

Dans la plupart des cas, le dérivé appartient à une catégorie
grammaticale différente de celle du mot originel. De ce fait, les
relations syntaxiques que le mot originel entretient avec son
environnement ne peuvent rester inchangées lors de la dérivation.

En examinant chaque type de dérivation suffixale, nous avons
construit une table de correspondance indiquant le schéma
syntaxique correspondant à chaque contexte originel. Ce tableau de
correspondance permet de transformer un fragment de texte par
dérivation d'une unité lexicale sans en modifier notablement le
sens. Par exemple, l'énoncé l'\textit{arrivée du train} peut être
transformé par un dérivé \textit{arriver} grâce au schéma
syntaxique correspondant (\texttt{NMOD[INDIR](arrivée,de,train)
==> SUBJ(arriver,train)}). L'énoncé qui en découle \textit{le
train arrive} a pratiquement la même signification que celui dont
il découle.

\section{Interroger la structure informationnelle}

Même si la base documentaire a été analysée et qu'une expansion
l'a considérablement enrichie, la phase d'analyse de la requête
seule permet d'identifier les données proches de la réponse à la
question. Le traitement de la requête est donc de première
importance. Ce traitement est similaire à celui des documents,
mais il est limité à l'analyse linguistique, et encore cette
dernière ne comprend-elle pas la désambiguïsation sémantique. La
procédure d'expansion n'est pas nécessaire car la structure
informationnelle a déjà été enrichie, et de plus elle ne peut être
mise en {\oe}uvre par manque de contexte permettant l'application
de la désambiguïsation sémantique.

L'analyse linguistique de la question amène à la construction
d'une structure informationnelle locale, similaire à celle des
documents, mais évidemment plus rudimentaire. La grammaire
employée pour la question a été modifiée de telle manière que les
caractéristiques spécifiques à l'interrogation n'apparaissent pas
dans l'analyse, et donc pas non plus dans la structure locale. Si
l'interrogation porte sur un mot explicite de la question, ce mot
(appelé \textit{focus}) disparaît de la structure locale, mais ses
caractéristiques sémantiques sont maintenues sous forme de traits,
et son requises dans la réponse. Dans les autres cas, les traits
sémantiques du \textit{focus} dépendent de l'interrogatif.

Une fois la structure locale élaborée, la réponse peut être
trouvée dans la structure informationnelle documentaire par simple
comparaison avec cette structure locale. Le système peut être
interrogé à différents niveaux (texte, paragraphe, phrase), mais
dans le cas d'une application de question-réponse, seul le niveau
de la phrase est pertinent. Une réponse congrue est une phrase qui
contient les informations contenues dans la structure locale,
ainsi que l'élément non identifié qui correspond aux traits
sémantiques du \textit{focus}.

\begin{figure}[!h]
\begin{center}
\mbox{\parbox{10cm}{
\textsf{Question proposée~:}\\
De quel chef Domitien fut-il le successeur~?\\

\textsf{Structure de la question (dépendances)~:}\\
\textbf{NMOD}[SPRED](\textbf{Domitien,successeur})\\
\textbf{NMOD[INDIR](successeur,de,chef)}\\

\textsf{Réponse attendue~:}\\
(\dots) Domitien succéda à l'empereur Titus (\dots)\\

\textsf{Structure enrichie du texte~:}\\
SUBJ(succéda \textit{OU remplacer},Domitien)\\
VARG[INDIR](succéda,à,empereur)\\
\textit{VARG[DIR](remplacer},empereur OU \textit{chef)}\\
NN(empereur OU \textit{chef OU souverain},Titus)\\
\textit{\textbf{NMOD[INDIR](successeur,de}},empereur \textit{OU \textbf{chef})}\\
\textit{\textbf{NMOD(Domitien,successeur)}}
}}
\caption{Interrogation d'un texte enrichi.} \label{interro}
\end{center}
\end{figure}

L'exemple \ref{interro} page \pageref{interro} montre comment
l'expansion du texte permet de mettre en correspondance une
question avec la réponse dans laquelle la plupart des termes de la
requête ne sont pas représentés. Les dépendances et leurs
arguments forment la structure informationnelle. Les parties en
gras montrent les éléments qui se répondent dans la structure de
la question et dans celle du texte. Les éléments en italiques sont
issus de l'enrichissement, soit synonymique, soit dérivationnel.

\section{Évaluation}

Étant donné qu'il n'existe aucune évaluation standard de la tâche
de question-réponse pour le français, nous nous sommes inspiré du
modèle de TREC-8 \cite{Voorhees99}. Pour notre évaluation, huit
personnes non spécialistes on lu 50 articles de
l'\textit{Encyclopédie Hachette Multimédia 2000} et on proposé 25
questions chacun sur leur lecture. Les 200 questions sont en
français correct, sans erreur orthographique ni syntaxique. Chaque
question possède au moins une réponse dans les articles. Nous
avons établi un <<~plancher~>> (\textsl{baseline}) en effectuant
une interrogation de la base documentaire en utilisant uniquement
le vocabulaire (lemmatisé) présent dans les documents, sans autre
traitement. Nous établissons un second point de comparaison en
enrichissant les documents avec tous les synonymes sans aucune
utilisation de l'analyse linguistique, et sans filtrage par la
désambiguïsation sémantique (\textit{synonymes sans sémantique}).

Nous comparons les résultats avec l'enrichissement sémantique
synonymique (\textit{synonymes avec sémantique}), puis avec tous
les enrichissements (\textit{tous les enrichissements}). Seules
les cinq premières réponses sont prises en compte. Une score est
alloué à chaque question. Ce score est de 1 si la première réponse
est correcte, de 1/2 si la première réponse est incorrecte et la
deuxième est correcte, de 1/3 si les deux premières réponses sont
fausses et que la troisième est correcte, etc. Si aucune bonne
réponse n'apparaît parmi les cinq premières, le score est de 0.
Les scores indiqués dans le tableau d'évaluation sont une moyenne
pour les 200 questions proposées.

\begin{table}[!h]
\begin{center}
\begin{tabular}{|l|c|c|}
\hline
Enrichissement              & Score  & Pas de réponse \\
\hline
Plancher                    & 0.295  & 139 \\
Synonymes (sans sémantique) & 0.303  & 137 \\
Synonymes (avec sémantique) & 0.487  & 100 \\
Tous les enrichissements    & 0.504  & 97 \\
\hline
\end{tabular}
\end{center}
\end{table}

Cette évaluation montre la réelle amélioration de l'utilisation
d'une discrimination sémantique pour l'expansion d'énoncés
appliquée à la tâche de question-réponse. Ces enrichissements
permettent d'éliminer de fausses propositions de réponses et
d'obtenir plus rapidement la réponse juste à une question posée.

Malgré ces résultats prometteurs, de nombreuses améliorations
peuvent encore être effectuées pour améliorer les performances du
système. Notamment, le manque de synonymes est encore flagrant
lorsqu'on étudie les questions et les (manques de) réponses en
profondeur. Par exemple, \textit{bru} et \textit{belle-fille} ne
sont jamais mis en relation. D'autres relations sémantiques
devraient également être intégrées pour permettre de faire le lien
entre des informations proches, mais pas synonymes~: l'hypéronymie
permettrait ainsi de mettre en relation \textit{fonction} et
\textit{consul}, le premier recouvrant le second. Il y a encore la
nécessité de marquer l'équivalence entre certaines expressions,
comme \textit{X est fatal à Y} qui équivaut à \textit{Y meurt de
X}.

Par ailleurs, l'exigence d'une correspondance stricte et exacte
entre la structure de la question et celle des documents elimine
de fait des réponses légèrement différentes de la structure
exigée. Une méthode de dégradation de la correspondance des
structures, avec pondération des réponses candidates, serait sans
doute la bienvenue.

\section{Conclusion}

Dans cet article, nous avons montré l'intérêt du prétraitement des
documents contenus dans une base documentaire destinée à être
interrogée selon les modalités de la tâche de question-réponse. Le
fait de soumettre le texte à des traitements plutôt que la
question est justifié par le contexte généralement plus large dans
les énoncés des documents, rendant de ce fait possible l'usage
d'un système de désambiguïsation sémantique pour identifier le
sens correct de chaque élément d'information dans le contexte dans
lequel il apparaît. De ce fait, les différents enrichissements
apportés pour réaliser une expansion de texte sont plus précis et
les réponses apportées sont plus souvent correctes.

Nous présentons également un système de question-réponse fondé sur
la construction d'une structure informationnelle d'une base
documentaire. Cette structure provient de l'analyse linguistique
(morphologique, syntaxique et sémantique) des documents, et de
l'expansion des énoncés par enrichissement synonymique et
dérivationnel. L'évaluation confirme l'intérêt de l'enrichissement
sémantique, qui est rendu possible seulement par le traitement des
textes plutôt que celui des questions.


\begin{thebibliography}{9}

\bibitem[Aït-Mokhtar, 1998]{Ait98}
Salah Aït-Mokhtar, \textit{L'analyse présyntaxique en une seule
étape}, Thèse de doctorat, Université Clermont 2 Blaise Pascal,
Clermont-Ferrand, 1998.

\bibitem[Aït-Mokhtar et al., 2002]{AitAl02}
Salah Aït-Mokhtar, Jean-Pierre Chanod, et Claude Roux, "Robustness
beyond shallowness: incremental deep parsing", dans
\textit{Natural Language Engineering}, tm. 8(2/3), pp. 121-144,
2002.

\bibitem[Bailly, 1947]{Bailly47}
René Bailly, \textit{Dictionnaire des synonymes de la langue
française}, Larousse, Paris, 1947.

\bibitem[Brun et al., 2001]{BrunAl01}
Caroline Brun, Bernard Jacquemin et Frédérique Segond,
"Exploitation de dictionnaires électroniques pour la
désambiguïsation sémantique lexicale", dans \textit{Traitement
Automatique des Langues}, tm. 42(3), pp. 667-690, 2001.

\bibitem[Corréard and Grundy, 1994]{CorreardGrundy94}
Marie-Hélène Corréard and Valerie Grundy (réds.), \textit{The
Oxford-Hachette French Dictionary}, Oxford University Press,
Hachette, Oxford, Paris, 1994.

\bibitem[Dubois and Dubois-Charlier, 1997]{DuboisDuboisCharlier97}
Jean Dubois et Françoise Dubois-Charlier, \textit{Dictionnaire des
verbes français}, Larousse, Paris, 1997. Ce dictionnaire n'existe
au départ qu'en version électronique. Il dispose de son complément
naturel \textit{Dictinnaire des mots}. Ces deux ressources sont
disponibles à ELDA.

\bibitem[Ferret et al., 1999]{FerretAl02}
Olivier Ferret, Brigitte Grau, Martine Hurault-Plantet, Gabriel
Illouz, et Christian Jacquemin, "Quand la réponse se trouve dans
un grand corpus", dans \textit{Revue d'Ingénierie des Systèmes
d'Information}, tm. 7(1-2), pp. 95-123, 2002.

\bibitem[Hagège and Roux, 2002]{HagegeRoux02}
Caroline Hagège et Claude Roux, "A Robust and Flexible Platform
for Dependency Extraction", in \textit{Proceedings of LREC 2002},
Las Palmas, Canaria, Espa\~{n}a, pp. 520-523, 2002.

\bibitem[Harabagiu et al., 2002]{HarabagiuAl02}
Sanda Harabagiu, Dan Moldovan, Marius Psca, Rada Mihalcea, Mihai
Surdeanu, Razvan Bunescu, Roxana Gîrju, Vasile Rus et Paul
Morarescu, "Falcon: Boosting Knowledge for Answer Engines", dans
2002.

\bibitem[Hull, 1999]{Hull99}
David A. Hull, "Xerox TREC-8 Question Answering Track Report",
dans Ellen M.  Voorhees et Donna Harman (réds.),
\textit{Proceedings of The Eighth Text Retrieval Conference
(TREC-8)}, pp. 743-752, 1999.

\bibitem[Jacquemin, 2003]{Jacquemin03}
Bernard Jacquemin, \textit{Construction et interrogation de la
structure informa- tionnelle d'une base documentaire en français},
thèse de doctorat, Université Paris III Sorbonne Nouvelle, Paris,
2003.

\bibitem[Jacquemin et al., 2002]{JacqueminAl02}
Bernard Jacquemin, Caroline Brun et Claude Roux, "Enriching a text
by semantic disambiguation for information extraction", dans
Claude de Loupy (réd.), \textit{LREC 2002 Workshop Proceedings.
Using Semantics for Information Retrieval and Filtering}, Las
Palmas, Canaria, Espa\~{n}a, 2002.

\bibitem[Roux, 1999]{Roux99}
Claude Roux, "Phrase-driven parser", dans \textit{Proceedings of
VEXTAL'99}, Venezia, Italia, pp. 235-240, 1999.

\bibitem[Segond et al., 2000]{SegondAl00}
Frédérique Segond, \'{E}lisabeth Aimelet, Veronika Lux  et Corinne
Jean, "Dictionary-driven Semantic Look-up", dans \textit{Computer
and the Humanities. Special Issue on SENSEVAL}, tm. 34(1-2), pp.
193-197, 2000.

\bibitem[Trouilleux, 2001]{Trouilleux01}
François Trouilleux, \textit{Identification des reprises et
interprétation automatique des expressions pronominales dans des
textes en français}, thèse de doctorat, Université Clermont 2
Blaise Pascal, Clermont-Ferrand, 2001.

\bibitem[Voorhees, 1999]{Voorhees99}
Ellen M. Voorhees, "The TREC-8 Question Answering Track Report",
dans Ellen M.  Voorhees et Donna Harman (réds.),
\textit{Proceeding of the Eighth Text REtrieval Conference
(TREC-8)}, pp. 77-82, 1999.

\bibitem[Voorhees et Harman, 1999]{VoorheesHarman99}
Ellen M.  Voorhees et Donna Harman (réds.), \textit{Proceeding of
the Eighth Text REtrieval Conference (TREC-8)}, 1999.

\end{thebibliography}
\end{document}